\documentclass[doublecol]{epl2} 
\usepackage{graphicx}
\usepackage{amsmath}
\usepackage{amssymb}

\title{Species separation and modification of neutron diagnostics in inertial-confinement fusion}

\author{A. Inglebert \thanks{E-mail: \email{aurelie.inglebert@cea.fr}} \and B. Canaud \and O. Larroche}
\shortauthor{A. Inglebert \etal}
\institute{                    
CEA, DAM, DIF, 91297 Arpajon Cedex, France
}
\pacs{52.57.-z}{Laser inertial confinement}
\pacs{52.65.Ff}{Fokker-Planck and Vlasov equation}
\pacs{52.35.Tc}{Shock waves and discontinuities}

\abstract{The different behaviours of deuterium (D) and tritium (T) in the hot spot of marginally-igniting cryogenic DT inertial-confinement fusion (ICF) targets are investigated with an ion Fokker-Planck model. With respect to an equivalent single-species model, a higher density and a higher temperature are found for T in the stagnation phase of the target implosion. In addition, the stagnating hot spot is found to be less dense but hotter than in the single-species case. As a result, the fusion reaction yield in the hot spot is significantly increased. Fusion neutron diagnostics of the implosion find a larger ion temperature as deduced from DT reactions than from DD reactions, in good agreement with NIF experimental results. ICF target designs should thus definitely take ion-kinetic effects into account.}

\begin{document}
\maketitle

\section{Introduction}

In inertial-confinement fusion (ICF) \cite{NUC724,LIN951,ATZ042}, a capsule containing a cryogenic deuterium (D) and tritium (T) mix is compressed and heated by lasers to reach a pressure and a temperature high enough to generate fusion reactions. The laser energy is delivered to the capsule either by directly hitting the target shell, or after being first absorbed in an enclosing hohlraum (the so-called indirect drive scheme \cite{LIN951}). In the former case, an additional laser pulse can be used to launch a strong shock wave through the fuel to trigger ignition \cite{BET071,CAN10NJP}. In either case, ablation of the shell by the deposited energy leads to fuel compression through a series of shock waves propagating across the DT plasma.
The implosion velocity of the fuel increases until the shocks bounce off the center of the target. It then decreases until the stagnation phase where most of the nuclear reactions are produced.

In that process, it has recently been argued that a separation might occur between the ion species comprising the fuel, due to baro-, thermo-, and electrodiffusion \cite{AME105,KAG122}. Such a phenomenon has been put forward to explain the deficit in nuclear reactivity recorded in direct-drive implosions of targets containing an equimolar gaseous mix of D and $\mathrm{^3He}$ on \textsc{omega} \cite{RYG069}. The result was compared with the appropriately scaled value expected from pure-$\mathrm{D_2}$ implosions, and with 1-D hydrodynamics simulations where species separation could not be taken into account. However, the ion-kinetic simulations that we conducted \cite{LAR12D} did not confirm that explanation, spawning further discussion\cite{AME115,BEL138,LAR131}.

The question of possible ion species separation during the deceleration phase of cryogenic DT targets is still of concern in the pursuit of ignition on the NIF \cite{LIN14Z} (and, in the near future, the LMJ \cite{GIO069}). The kinetic description of the deceleration phase of indirect-drive high-gain target design was addressed a few years ago \cite{LAR03A}, however using an average ion species with an atomic mass $A=2.5\un{g/mol}$, as usually done in hydrodynamics simulations, which obviously can't address ion species separation.

\section{Ion-kinetic simulation of implosion}

This work addresses kinetic effects, including ion species separation, in the hot spot of a set of various targets, using the ion Fokker-Planck code \textsc{fpion} \cite{LAR03A,LAR931}. For the sake of definiteness, our calculations will be described in more detail in the following in the case of a direct-drive target designed \cite{BRA135} for shock \cite{BRA13WOC} or standard ignition \cite{BRA14NF}. It is a marginally igniting spherical capsule of cryogenic deuterium (D) and tritium (T) mixture enclosed in a CH ablator with a low aspect (radius to thickness) ratio, which is to be contrasted to the target studied in \cite{LAR03A}. At stagnation, the hot spot radius in \cite{LAR03A} is about 30~$\mu m$ with a thermal ion-ion collision mean free path around 17~nm, while in our case, the hot spot radius is about 25~$\mu m$ with a shorter collision mean free path (around 8~nm).

We first investigate the deceleration phase of the implosion where ion species separation could take place, beginning when its peak velocity reaches a maximum value of 290~km/s, 11.65~ns after the beginning of the implosion. We then examine the impact of that differential species behaviour on reactivities and neutron diagnostics.

The code considers a spherical geometry with one spatial dimension $r$ and two velocity dimensions $(v_r,v_\perp)$. Each ion species is characterized by a distribution function $f_i(r,v_r,v_\perp)$ governed by the Vlasov-Fokker-Planck equation:
\begin{align*}
\frac{\partial f_i}{\partial t}+ v_r \frac{\partial f_i}{\partial r}+&\frac{v_{\perp}}{r}\left ( v_{\perp}  \frac{\partial f_i}{\partial v_r}-v_r \frac{\partial f_i}{\partial v_{\perp}}\right)+\frac{E_i}{A_i}\frac{\partial f_i}{\partial v_r}  \\
& =\sum_{j=1}^{n} \left( \frac{\partial f_i}{\partial t} \right)_{ij}+\left(\frac{\partial f_i}{\partial t} \right)_{ie}
\end{align*}
where $E_i$ is the effective electric force applied to ions of species $i$ with  atomic mass $A_i$. The electric field is related to the electron pressure gradient $ E_i = -Z_i/n_e \partial P_e / \partial r$ (where $n_e$ and $P_e$ are the electron density and pressure respectively). The electric force is a source of ion species separation due to differences in their charge-to-mass ratio $Z_i/A_i$. The first term in the right-hand side accounts for ion-ion collisions, the second one  for ion-electron collisions. They are described in more detail in \cite{LAR03A,LAR931}.
Electrons are considered as a background neutralizing fluid of density $n_e$ and temperature $T_e$ governed by a thermal transport equation, with an equation of state taking quantum degeneracy into account \cite{LAR03A}.

Velocity, temperature, and density profiles are extracted from hydrodynamics calculations and remapped onto the \textsc{fpion} eulerian grid to provide an initial condition for the kinetic simulation. The outermost fuel mesh in the Lagrangian hydrodynamics simulation is used to create the boundary conditions for the Fokker-Planck calculation of the deceleration stage.

\section{Typical results}

\begin{figure}[t]
\centerline{\includegraphics[width=8.7cm]{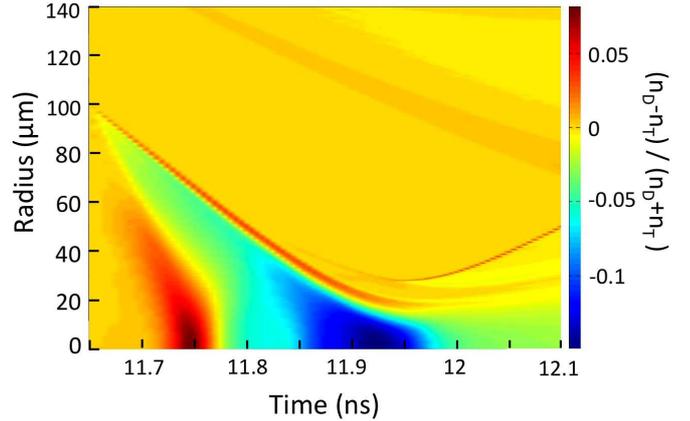}}
\caption{Variation of the relative density difference $(n_D-n_T)/(n_D+n_T)$ versus space $r$ ($\mu$m) and time $t$ (ns) during the whole deceleration phase inside the DT fuel. Two regions appear; the red one corresponds to an excess of deuterium and a lack of tritium, the blue one to a predominance of tritium.
}
\label{dn}
\end{figure}

To evidence species separation, we plot on fig.~\ref{dn} the temporal and spatial variations of the relative abundance ${(n_D-n_T)/(n_D+n_T)}$ during deceleration (beginning at $t=11.65\un{ns}$), where $n_D$ and $n_T$ are the deuterium and tritium densities, respectively.

Different regions can be observed near fuel center. The first one (in red), between $11.7\un{ns}$ and $11.8\un{ns}$, corresponds to the shock collapse at the target center with a 7\% relative excess of deuterium over tritium. Then, the shock bounces back outwards, and the second part (in blue) displays a tritium excess, with a relative density higher by roughly 15\%. This separation occurs in the hot-spot; the thin ``red strip'' in the figure is close to the boundary between the hot spot and the denser surrounding main fuel.

Let us mention that new calculations of the indirect drive target studied in \cite{LAR03A}, but now modeling the DT fuel with two different ion species, also exhibited ion separation, with relative density variations of the same order of magnitude.

We can also notice that some ion species separation occurs in the inner part of the dense fuel, as evidenced by the thin red strip in fig.~\ref{dn}. Since pressure varies smoothly with radius in that region, ion separation arises from temperature gradients at the hot spot/dense fuel boundary, leading to a deuterium-enriched layer in the inner part of the shell and a lack of deuterium in the outer layer of the hot spot, much as discussed in \cite{MOL120}.

\begin{figure}[t]
\centerline{\includegraphics[width=8.7cm]{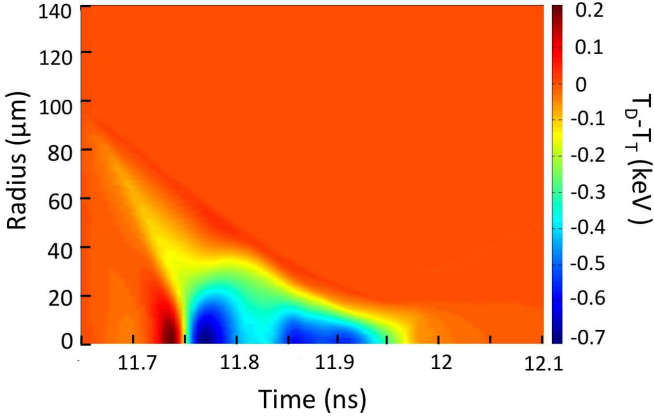}}
\caption{Difference of ion temperatures $T_D-T_T$, in keV, versus space $r$ ($\mu$m) and time $t$ (ns) during the whole deceleration phase. Two regions appear: the red one corresponds to a deuterium temperature higher than tritium ($T_D>T_T$) and the blue one is the opposite ($T_D<T_T$). }
\label{dT}
\end{figure}

Moreover, just before stagnation, tritium and deuterium have different temperatures, as shown on fig.~\ref{dT}. Tritium is about $700\un{eV}$ hotter than deuterium when its abundance is higher. When D is predominant, its temperature is higher by $150\un{eV}$.

\begin{figure}[t]
\centerline{\includegraphics[width=8.7cm]{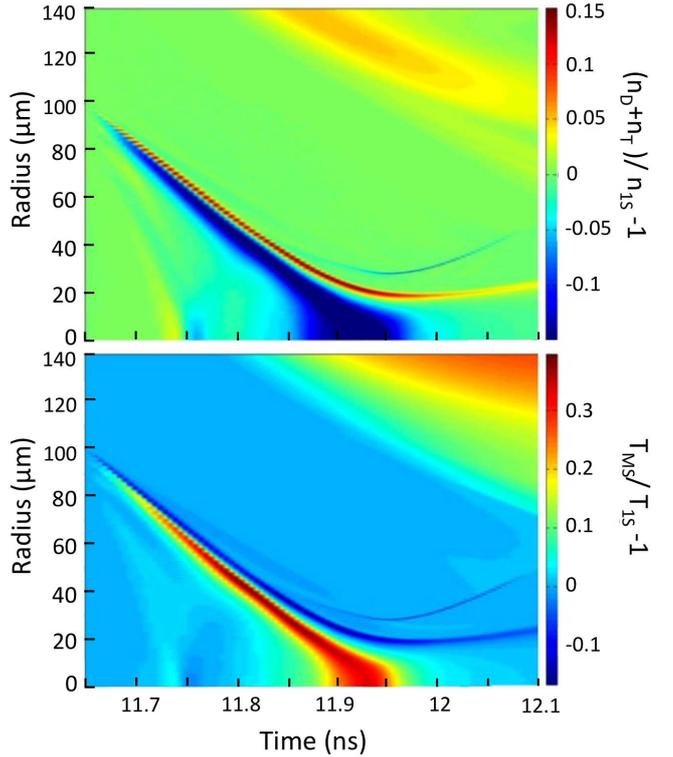}}
\caption{Density (top) and temperature (bottom) relative differences $n_{MS}/n_{1S}-1$, $T_{MS} /T_{1S}-1$ between a single-species plasma ($A=2.5\un{g/mol}$) and the two-ion-species DT case. At $t\simeq 11.94\un{ns}$, the two-species case is about 15\% less dense and 35\% hotter than the single-species plasma.}
\label{comp_mono}
\end{figure}

A direct comparison with 1D hydrodynamics calculations is not straightforward, due to numerous differences between physical descriptions and assumptions in each model. In order to focus on the effects of ion species separation, we performed kinetic calculations considering only a single species, modeling a DT equimolar mixture with an average atomic mass $A=2.5\un{g/mol}$.
All other parameters and initial and boundary conditions remained identical to the multispecies (MS) calculations. The single-species density $n_{1S}$ and temperature $T_{1S}$ are compared with the two-ion-species density $n_{MS}=n_D+n_T$ and average temperature $T_{MS}=\left(n_DT_D+n_TT_T\right)/(n_D+n_T)$.
The relative differences in density $n_{MS}/n_{1S}-1$ and temperature $T_{MS}/T_{1S}-1$ are plotted as functions of space and time on fig.~\ref{comp_mono}.

We checked that there is no temporal and space differences between both cases.  Early in the deceleration phase ($t\sim 11.7\un{ns}$), the deuterium excess leads to a 3\% increase of the average ion density in spite of the lack of tritium. Hence the mass density also displays a 12\% excess. After the first shock convergence at the center, deuterium propagates back toward the dense fuel, leading to a lack of deuterium at stagnation in the hot-spot.  Tritium becomes predominant, but the average density is 15\% lower than in the single-species case, resulting in a 30\% lower mass density. At convergence time, multi-species ion temperatures don't significantly depart from the single-species one. On the contrary, around stagnation time, when deuterium is depleted, the multi-species plasma is 35\% hotter.

\section{Consequences on fusion reactions}

\begin{figure}[t]
\centerline{\includegraphics[scale=0.15]{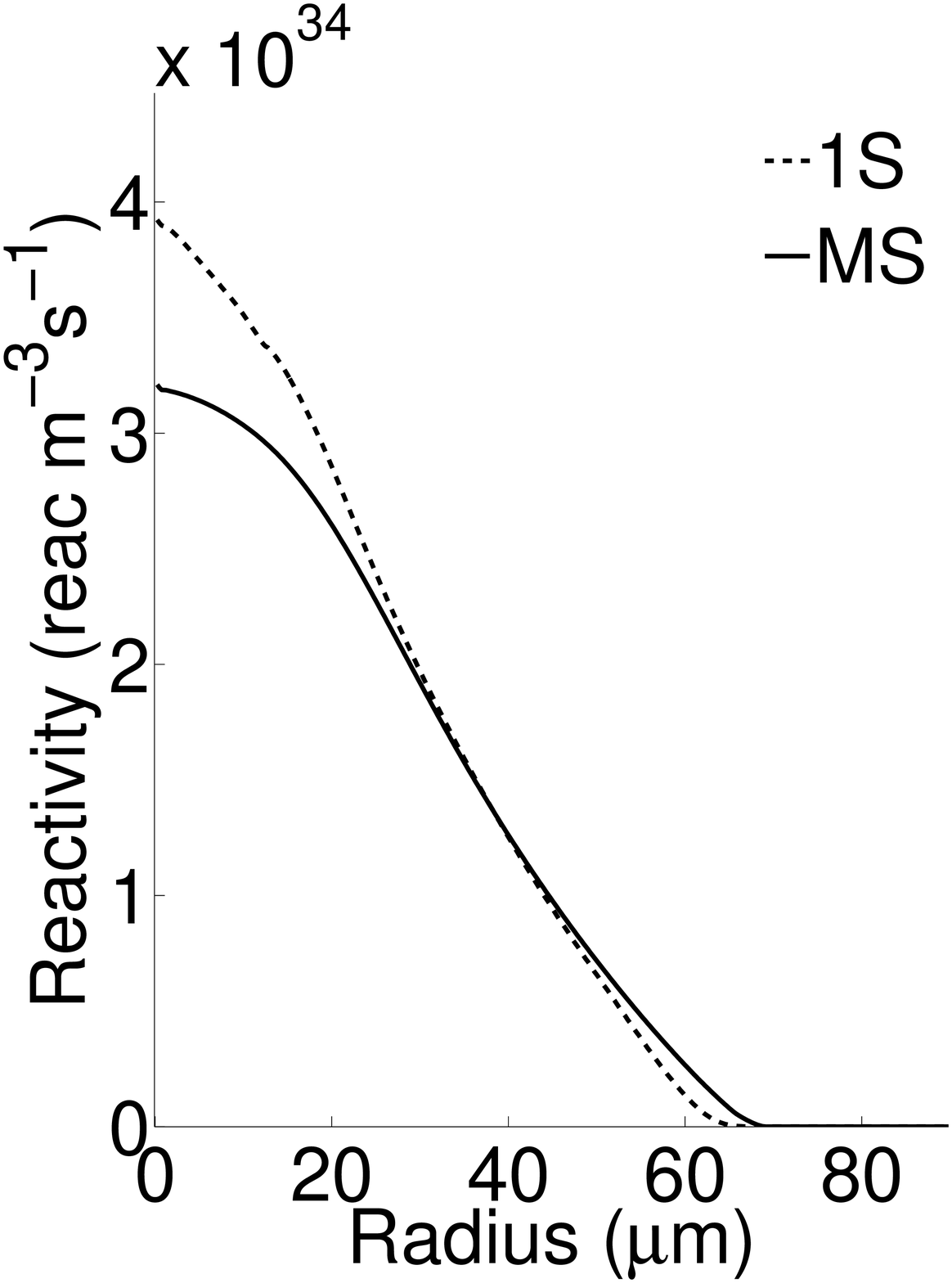}
\includegraphics[scale=0.15]{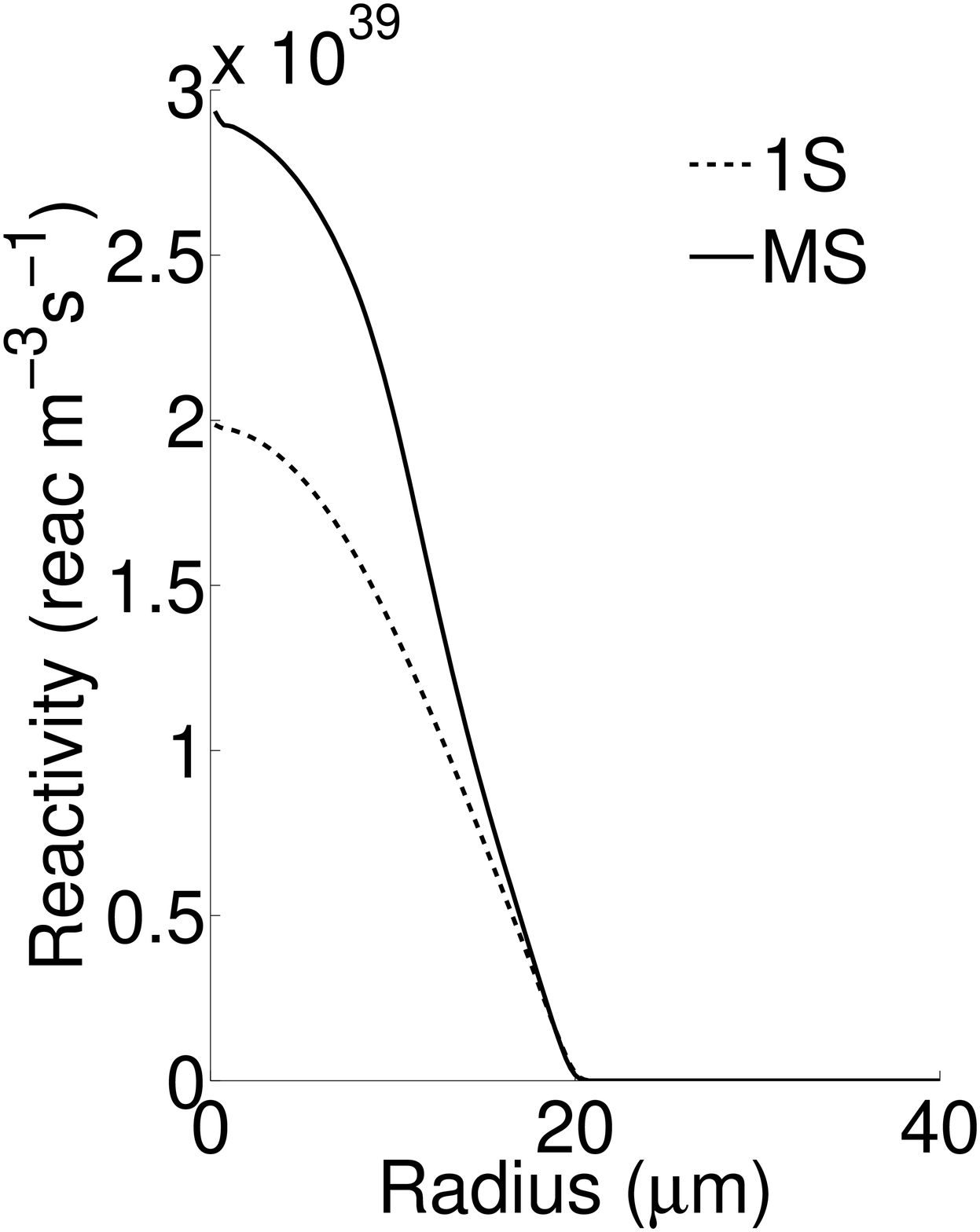}}
\caption{Reactivity for the D+T reaction at $t=11.75\un{ns}$ (left) and $t=11.94\un{ns}$ (right) in and around the hot-spot. The solid lines correspond to the two-ion-species case, and the dotted lines to the single-species equimolar case with an average atomic mass $A=2.5\un{g/mol}$.}
\label{reacDT}
\end{figure}

These density and temperature differences will have consequences on neutron production and could modify the neutron yield predicted from the hot spot of ICF targets. They might also affect measurements \cite{RIZ064,HOU06B} from neutron-based diagnostics. Thermonuclear fusion reactions are computed in our code as a post-processing diagnostic from distribution functions:
\begin{equation}
\begin{aligned}
R_{ij}=\frac{1}{1+\delta_{ij}} \int f_i(\vect{v_i}) &f_j(\vect{v_j}) \times\\
 |\vect{v_i}-\vect{v_j}|\sigma_{ij}&(|\vect{v_i}-\vect{v_j}|)\upd^3v_i\upd^3v_j \label{equation_reac}\end{aligned}
\end{equation}
where indices $i$ and $j$ refer to ion species D or T, $\delta_{ij}$ is the Kronecker delta and $\sigma_{ij}(|\mathbf{v_i}-\mathbf{v_j}|)$ is the velocity-dependent cross-section for the $i+j$ reaction, estimated from the parametrization of \cite{BOS922}. We consider the following two fusion processes:
\begin{align*}
& \chem{D}+\chem{T} \rightarrow \chem{^4He}+\chem{n} \, (14.1\un{MeV})\\
& \chem{D}+\chem{D}\ ^{\underrightarrow{\, 50\% \, }}\ \chem{^3He}+\chem{n} \, (2.45\un{MeV})
\end{align*}
Figure~\ref{reacDT} displays the D+T reactivity as a function of radius inside the target at two different times. Solid lines correspond to the two-ion-species case, and dotted lines to the single-species one.
The first time $t=11.75\un{ns}$ is when species separation first occurs in the two-species case, with an excess of deuterium over tritium (cf.~fig.~\ref{dn}). At that time, as shown on fig.~\ref{reacDT}, the reactivity of the multi-species plasma, around $3.2\times10^{34}\un{reactions.m^{-3}.s^{-1}}$, is lower than in the single-species case (with $\simeq 4\times10^{34}\un{reactions.m^{-3}.s^{-1}}$), due to this lack of tritium. 
At stagnation, when tritium accumulates around the fuel center (at $t\simeq11.94\un{ns}$), the multi-species plasma is less dense but warmer than the single-species plasma. That difference directly affects reactivity which, in spite of species separation, ends up larger for the two-species case ($\simeq 3\times 10^{39}\un{m^{-3}.s^{-1}}$) than in the single-species plasma ($\simeq 2\times10^{39}\un{m^{-3}.s^{-1}}$). This difference is mainly due to the increase of the tritium temperature at stagnation.
It is worth noticing that the yield at stagnation is five orders of magnitude higher than at shock convergence time, when deuterium is predominant. Species separation in that first stage is thus not expected to have any effect on the neutron yield ultimately recorded.

We also see that kinetic effects in the vicinity of the hot spot/dense fuel interface (the ``Knudsen layer'' effect \cite{MOL120}) are not significant since the reactivity there is negligible as compared with the hot-spot one.

\begin{figure}[t]
\centerline{\includegraphics[scale=0.19]{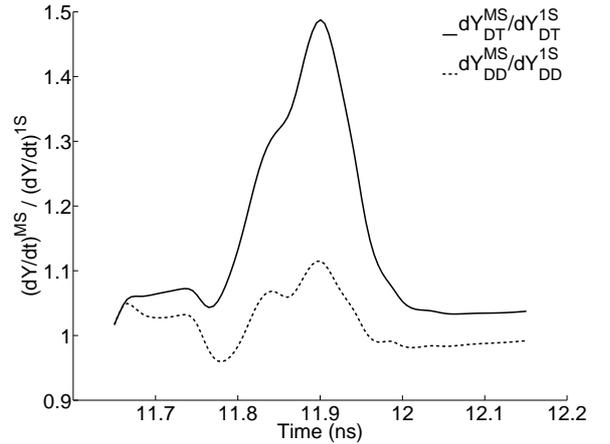}}
\caption{Ratios between the multi-species (MS) and single-species (1S) values of the time derivative of the D+T (solid line) and D+D (dotted line) yields as functions of time. More DT and DD reactions per unit time are produced in the two-ion-species case than in the single-species one. }
\label{yield dt}
\end{figure}

By integrating these reactivities over space, we calculate the target neutron yield as a function of time $dY/dt$, and then in turn the total number of neutrons produced during the whole deceleration phase.
Figure \ref{yield dt} displays the ratio of the multi-species (MS) to single-species (1S) values of the time derivative of the yields, for the D+T (solid line) and D+D (dotted line) reactions as functions of time: $(dY/dt)^{MS}/(dY/dt)^{1S}$. 
In the single-species case, the reactivities calculated by \textsc{fpion} are obtained from eq.~(\ref{equation_reac}) using Maxwellian distributions based on first moments of the kinetic single-species distribution.
The reaction rates for D+T and D+D are greater in the multi-species case than in the single-species one. The ratio MS/1S first displays a small bump when deuterium is predominant.
Then, just before stagnation, a large peak appears, corresponding to tritium predominance and due to the average temperature increase in the MS plasma in spite of a reduced average ion density (fig.~\ref{comp_mono}). Thus, species separation in the multi-species case is seen to improve yields at the end of the deceleration phase. This reactivity increase just before stagnation should be beneficial to shock ignition.

\section{Neutron diagnostics of the implosion}

\begin{figure}[t]
\centerline{\includegraphics[scale=0.18]{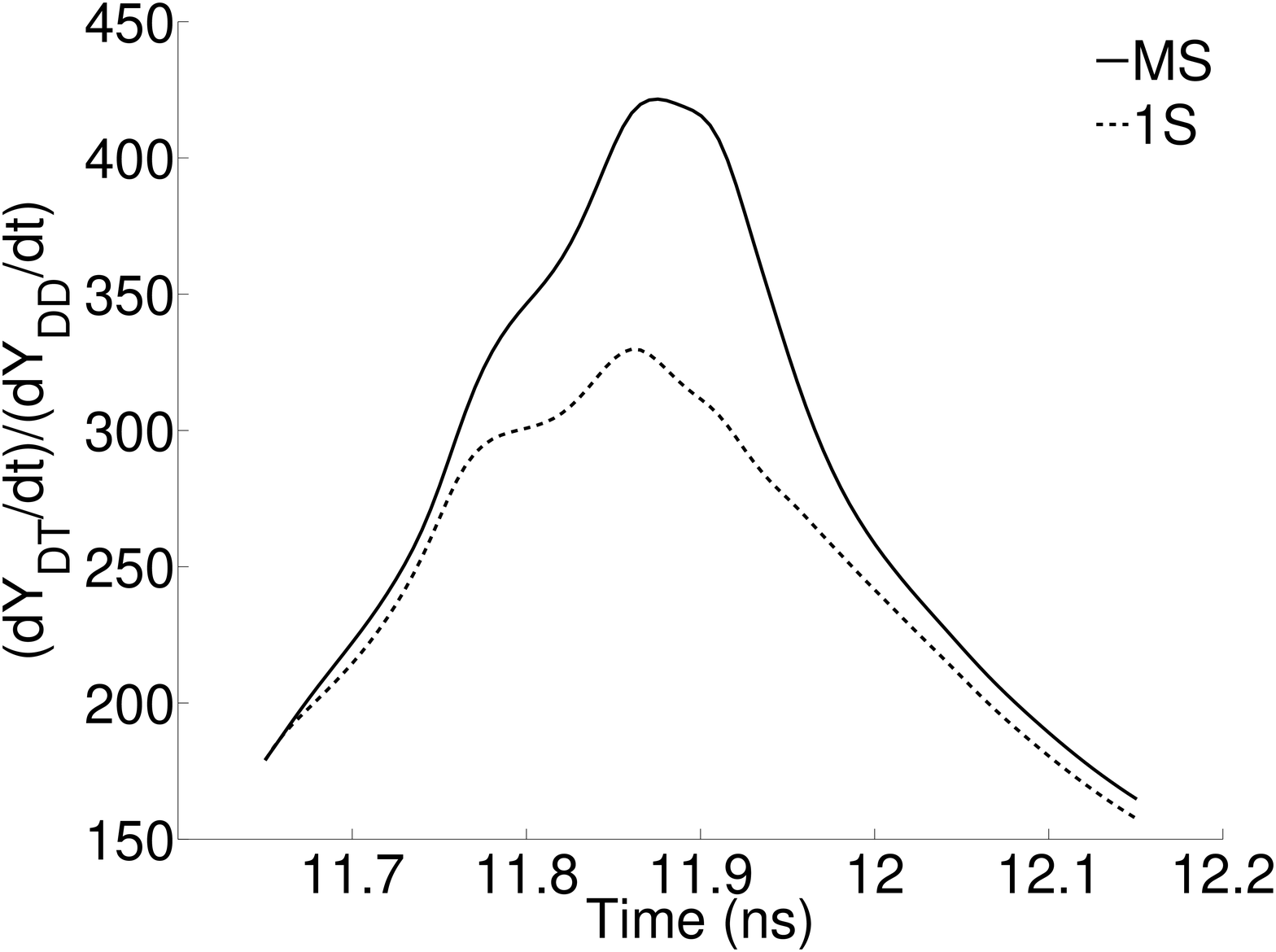}}
\caption{Ratios between the D+T and D+D reactivities as functions of time for the two-ion-species case (solid line) and the single-species one (dotted line). When species separation occurs, more reactions per unit time are produced for the two-ion-species case.}
\label{yield dtdd}
\end{figure}

In order to estimate a modification of the neutron spectrum, the ratio between the D+T and D+D reaction yields $(dY_{DT}/dt)/(dY_{DD}/dt)$ is plotted on Figure~\ref{yield dtdd} for the single-species and two-ion-species cases. The time derivative of the MS neutron yield is greater than the single-species one, due to the increased efficiency of the D+T reaction from the higher tritium temperature at stagnation, despite the decrease in average ion density. 

By integrating these results over time, we can calculate the overall yield ratio of the D+T and D+D reactions for both cases. For the multi-species mixture, we find a ratio $Y_{DT}/Y_{DD}=3.99\times10^{15}/1.34\times10^{13}=300$. For the single-species plasma, we calculate $Y_{DT}/Y_{DD}=3.41\times 10^{15}/1.36\times10^{13}=250$.
These estimates are source terms of the neutron emission.

\begin{figure}[t]
\centerline{\includegraphics[scale=0.6]{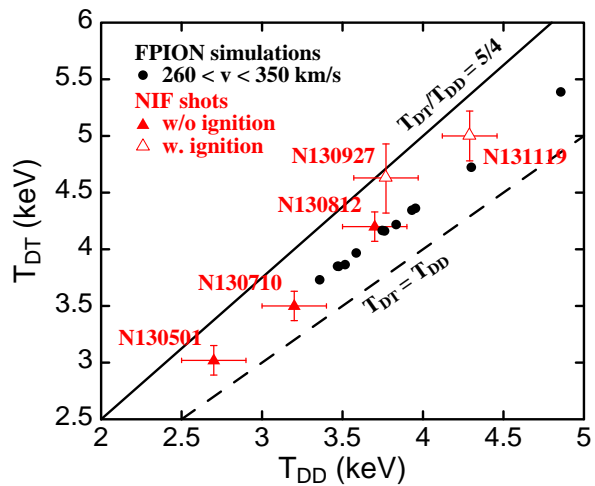}}
\caption{DT- versus DD-reconstructed ion temperatures for NIF shots \cite{HUR14M,PAR140} (triangles) and for \textsc{fpion} simulations (filled circles). A wide variety of target designs is considered here, varying the implosion velocity, the initial aspect ratio and the in-flight adiabat. The solid line (with slope 5/4) is the upper boundary of the region where targets with appreciable Doppler broadening from bulk hydrodynamic motion should be found.}
\label{fig3_T}
\end{figure}

In ICF experiments, the temperature of reacting ions is estimated from the measured energy spectrum of the fusion-produced neutrons, through a deconvolution process of the Doppler broadening effect \cite{BRY731,FRE135}. Two temperature values, $T_{DT}$ and $T_{DD}$, can thus be deduced from the 14 MeV DT and 2.45 MeV DD neutron emissions, respectively. Those values are computed from formula~(13) of ref.~\cite{FRE135}, which is itself a special case of formula~(10) of ref.~\cite{BRY731} when the D and T ion temperatures are equal. Going back to the latter, we see that what is actually measured through neutron spectra is a mass-weighted average of the reacting ion temperatures, namely:
\begin{equation}
\frac{m_DT_D+m_TT_T}{m_D+m_T} = T_{DT}\quad\mbox{and}\quad T_D = T_{DD} \label{eqTDTTDD}
\end{equation}
In order to model the experimental measurements of $T_{DT}$ and $T_{DD}$ in our kinetic calculations, we thus computed the mass-averaged temperature
\begin{equation}
T_{ij} = \frac{1}{Y_{ij}}\int \upd t\int \upd V \frac{(m_iT_i+m_jT_j)}{(m_i+m_j)}R_{ij} \label{eqTijFP}
\end{equation}
where $R_{ij}$ is the reaction rate from eq.~(\ref{equation_reac}), $m_i$ is the mass of ions of species $i$, $dV$ is the fuel volume element and
\begin{equation*}
Y_{ij} = \int R_{ij}\upd V\upd t
\end{equation*}
The quantities $T_{DT}$ and $T_{DD}$ were computed according to eq.~(\ref{eqTijFP}) in the case of the target described above as well as other targets with completely different implosion histories and parameters \cite{BRA135,CAN077,BRA14NF}. The latter include different implosion velocities, initial aspect ratios, target architectures, drivers, or numerical parameters. Figure~\ref{fig3_T} displays $T_{DT}$ as a function of $T_{DD}$ for those
\textsc{fpion} calculations (dots) as well as for experimental results recorded in recent NIF cryogenic implosions \cite{HUR14M,PAR140} (triangles). The first set of NIF shots \cite{PAR140} involves non-igniting cryogenic DT targets (shots number N130501, N130710, and N130812) while the other ones \cite{HUR14M} exhibits an alpha energy re-deposition in the plasma of the same order as mechanical ($PdV$) work.
As shown in fig. \ref{fig3_T}, $T_{DT}$ is systematically greater than $T_{DD}$,
in NIF experiments as well as in our calculations where a wide variety of target designs is treated. Up to now, the international community focussed on $T_{DT}$, implicitly assuming that $T_{DT}=T_{DD}$ in order to compare with and validate numerical hydrodynamics calculations. Those new and unexpected measurements can change our perception of the end-of-implosion state and consequently the predictions of ignition conditions. Indeed, dividing term by term the expressions in (\ref{eqTDTTDD}), the T/D relative temperature difference can be related to the measured relative differences:
\begin{equation}
\frac{T_{DT}}{T_{DD}} - 1 = \frac{m_T}{m_D+m_T}\left(\frac{T_{T}}{T_{D}} - 1\right) \label{eqTDTsTDD}
\end{equation}
From the latter expressions we find that the relative T/D temperature differences found in our simulations (of order $10-15$\%) account in a quite satisfactory way for the relative $T_{DT}/T_{DD}$ differences found in the neutron diagnostics interpretation, as displayed on fig.~\ref{fig3_T}.

NIF shots with noticeable gain (N130927 and N131119) are characterized by a contribution of $\alpha$-redeposition to the total yield. ``Bootstrapping'' of fusion exhibited in ref.~\cite{HUR14M} is seen as self-heating of $^4$He ions that ``run away'' and deposit their energy in the fuel. The fraction deposited in the hot spot is estimated in ref.~\cite{HUR14M} to be 75\% of the total $\alpha$ incoming energy, the rest being lost in the dense shell. As a result of the increased temperature and pressure, the reacting region will start expanding outwards, with consequences on the observed neutron spectrum. From eq.~(34) of ref.~\cite{BRY731}, an isotropic expansion velocity $V$ will bring a additional contribution to $T_{ij}$ proportional to $(m_i+m_j)V^2$, thus increasing the estimated temperature ratio $T_{DT}/T_{DD}$ up to 5/4, depending on the relative magnitude of the expansion velocity with respect to the thermal ion velocity. However, in its present state our code can't take $\alpha$-particle energy deposition into account, and thus can't describe this effect in a quantitative way. This will be addressed in future work \cite{PEI143}.

\section{Conclusion}

In summary, we  present a study of hot spot ion species separation during the deceleration phase of the implosion of marginally igniting ICF targets. Using the ion Fokker-Planck code \textsc{fpion}, we first show that species separation does take place during the whole deceleration phase, and until stagnation. An excess of deuterium prevails at shock convergence, and then a tritium predominance around stagnation. Some species separation is also observed in the inner part of the dense fuel with an excess of deuterium, however with no effect on fusion reactivity. Comparison with a single-species case, with an average species of atomic mass $A=2.5\un{g/mol}$, shows that the two-ion-species case is about 35\% hotter at stagnation than the single-species one. This difference directly affects the fusion reactivity. A calculation of the D+T and D+D reaction rates shows that species separation improves the D+T yield, mostly due to the higher tritium temperature at stagnation, and modifies the neutron yield ratio between the D+D and D+T reactions. The deleterious effects of a higher kinetic thermal loss to the colder fuel found in \cite{LAR03A} are seen to be more than compensated for by the higher tritium temperature and reactivity.

In addition, appropriately taking into account ion kinetics and its consequences on the ICF target implosion process leads us to a satisfactory interpretation of the discrepancies found in ion temperatures based on experimental neutron diagnostics. We conclude that these physical effects should definitely be included in the target optimization process aiming at ignition and gain in ICF, thus possibly alleviating the obstacles that prevented the recently completed NIC campaign \cite{LIN14Z} from reaching its intended goal. Such a modelling has been undertaken recently \cite{ROS14S} to account for the strong ion-kinetic effects encountered in gas-filled, exploding-pusher-type targets. Let us point out that the T/D temperature difference, leaving aside the species separation effect \textit{per se}, should be easily modelled in present-day Lagrangian hydrodynamics codes, by including an additional Lagrangian variable for the other ion species temperature, and accordingly an additional set of equations for the transport and relaxation of the new ion temperature variable.

\acknowledgments
One of the authors (BC) would like to thank L. Masse and P. Gauthier for fruitful discussions.

\end{document}